\documentclass[12pt,a4paper]{article}
\usepackage{amsmath,amssymb,epsfig,a4,latexsym}
\usepackage{physics}
\usepackage{slashed}
\usepackage{color}
\usepackage{rotating}
\usepackage{graphicx}
\usepackage{caption}
\usepackage[caption=false]{subfig}

\begin{document}
\thispagestyle{empty}
\begin{flushright}
PSI-PR-17-12\\
\today\\
\end{flushright}
\vspace{3em}
\begin{center}
{\large\bf The Standard Model and low-energy experiments: \\
from lepton-flavour violation to dark photons} 
\\ 
\vspace{3em}
{\sc G.~M.~Pruna\footnote{e-mail: Giovanni-Marco.Pruna@psi.ch}
}\\[2em]
{\sl Paul Scherrer Institut,\\
CH-5232 Villigen PSI, Switzerland}
\setcounter{footnote}{0}
\end{center}
\vspace{2ex}
\begin{abstract}
\noindent
This note is a concise theoretical summary of the first session on new physics searches at the high-intensity frontier of the IFAE2017 Conference.
Recent theoretical developments related to muonic lepton-flavour violation and $g-2$ are reviewed.
\end{abstract}
 
\newpage

\section*{Introduction}
\noindent
The Standard Model (SM) of particle physics with its most accurate description of fundamental interactions represents one of the greatest intellectual achievements of humankind. However, it does not satisfactorily explain the origin of matter, the nature of neutrino oscillations, the observation of dark matter and dark energy, and it does not accommodate gravity. Consequently, the SM is commonly accepted simply as a low-energy manifestation of an ultimate theory defined at the Planck energy scale, which should incorporate solutions to these open problems. The lack of evidence of any new physics (NP) signal from low- and high-energy experiments suggests that such NP completion could either be \emph{weakly coupled} to the SM or \emph{much heavier} than the electroweak (EW) symmetry-breaking scale (or a combination of both circumstances).

The experimental high-intensity frontier represents a very promising place to look for such NP scenarios: on the one hand, if beyond the SM (BSM) physics is realised at higher scales, then it is possible to search for deviations from the expected SM interactions (typically adopting an effective field theory (EFT) approach); on the other hand, lighter dark sectors can be investigated in searches for invisible decays and displaced interaction points.

During the first session on NP searches at the high-intensity frontier of the IFAE2017 Conference, the current~\cite{TheMEG:2016wtm} and future~\cite{Baldini:2013ke} experimental searches on muonic lepton-flavour violation, the status of the Muon $(g-2)$ experiment~\cite{Grange:2015fou}, the searches for invisible and/or exotic particles~\cite{Peccei:2006as,Dobrescu:2014fca,Dobrescu:2014ita,Flacke:2016szy} at BaBar~\cite{Lees:2017lec}, NA62~\cite{Lazzeroni:2017fza} and PADME~\cite{Raggi:2014zpa} were discussed. In what follows, we will focus on muonic experiments and briefly review some literature on the topic.

\section*{Testing the muonic sector of the SM}
\subsection*{Muonic lepton-flavour violation}
\noindent
To test efficiently for the presence of NP interactions, the experimental community has devoted a large effort to investigating processes that are forbidden/suppressed in the SM, due to the advantageous absence/suppression of background. One such promising study is the test of muonic lepton-flavour violating (LFV) decays, \emph{e.g.} $\mu\to e\gamma$~\cite{TheMEG:2016wtm}, $\mu\to 3e$~\cite{Bellgardt:1987du} and coherent $\mu\to e$ conversion in nuclei~\cite{Bertl:2006up}.

In addition, systematic efforts have also been devoted to understanding the theoretical aspects related to the interpretation of the absence of charged LFV (cLFV) signals in terms of limits on BSM physics. Background calculations were carried out for the so-called ``radiative''~\cite{Fael:2015gua,Fael:2016hnz,Pruna:2017upz} and ``rare''~\cite{Flores-Tlalpa:2015vga,Fael:2016yle,Pruna:2016spf} decays, \emph{i.e.} for the processes $l\to l'\gamma+2\nu$ and $l\to 3l'+2\nu$ (or $l\to l'+2l''+2\nu$), respectively.
Both these channels are important for the determination of the limits on the branching ratios of the two cLFV processes $l\to l'\gamma$ and $l\to 3l'$ (or $l\to l'+2l''$) because they provide an identical signal in the circumstance where invisible energy tends to zero, especially in view of the new experimental plans to improve the exploring power on these channels by MEG~II~\cite{Baldini:2013ke} and Mu3e~\cite{Blondel:2013ia}.
As for the phenomenological interpretation of the absence of a signal in terms of limits on the parameter space of potential BSM scenarios, a systematic EFT treatment was proposed~\cite{Pruna:2014asa,Pruna:2015jhf} and further developed with a particular focus on the muonic coherent LFV transitions~\cite{Crivellin:2016ebg,Crivellin:2017rmk}.
\vspace{-0.3cm}
\subsection*{The Muon $g-2$ experiment}
\noindent
A vast theoretical literature exists on the muonic anomalous magnetic moment $a_\mu$ and the current tension of $\sim 3$ standard deviations between the value obtained by the E821 experiment performed at the Brookhaven National Laboratory~\cite{Bennett:2006fi} and the theoretical prediction~\cite{Hagiwara:2011af}, and reviewing it in detail goes beyond the scope of this note. Instead, the attention of the reader is brought to the latest contributions and new phenomenological developments. Concerning the former, we recall that the relevant contributions can be classified in the following way~\cite{Hagiwara:2017zod}: quantum electrodynamics contributions~\cite{Aoyama:2012wk}, EW contributions~\cite{Gnendiger:2013pva}, hadronic light-by-light leading~\cite{Jegerlehner:2009ry} and next-to-leading~\cite{Colangelo:2014qya} order contributions, and hadronic vacuum polarisation leading~\cite{Davier:2010nc}, next-to-leading (see~\cite{Jegerlehner:2009ry} for a detailed review) and next-to-next-to-leading~\cite{Kurz:2014wya} order contributions.

Recently, a novel proposal~\cite{Calame:2015fva,Abbiendi:2016xup} suggested extracting the hadronic vacuum polarisation from the contribution to the muonic $g-2$ by exploiting $\mu e$ scattering.


{\par\medskip{\centering$\ast\ \ast\ \ast$\par}\medskip}
\let\stars\acknowledgments
\let\solong\acknowledgments
\noindent
This work was supported by the SNSF under contract 200021\_160156.


\end{document}